\documentstyle[12pt,psfig]{article}
\begin{document}

\author{
{\bf G. Herrera }
{\bf A. S\'anchez-Hern\'andez}\\
{{\small Centro de Investigaci\'on y de Estudios Avanzados}}\\
{{\small Apdo. Postal 14 740, M\'exico 07000, D.F.}}\\ \\
{\bf E. Cuautle},
{\bf J. Magnin}\\
{{\small Centro Brasileiro de Pesquisas F\'{\i}sicas}} \\
{{\small Rua Dr. Xavier Sigaud 150}}\\
{{\small 22290-180, Rio de Janeiro, Brazil}}
}

\title{
Production asymmetry of D mesons in $\gamma p$ collisions
\thanks{This work was supported by: 
CLAF (Brazil) and  
CONACyT (M\'exico), CIEA-JIRA (M\'exico)}
}

\date{}
\maketitle

\begin{abstract}
We study the production asymmetry of charm versus anticharm mesons
in photon-proton interactions. We consider photon gluon fusion plus
higher order corrections in which light quarks through vector meson 
-proton interactions contribute to the cross section. 
Non perturbative effects are   included in terms of a recombination 
mechanism which gives rise to a production asymmetry.
\end{abstract}

\newpage

\section{Introduction}

In high energy photon-hadron interactions, charm production is 
expected to be dominantly produced by photon-gluon fusion processes.
According with QCD perturbative calculations, this mechanism
produces in equal amounts  charm and anti-charm.\\
However, recent measurements of charm meson production
\cite{e687} indicate that there are important nonperturbative 
QCD phenomena in the production process that induce an asymmetry
in charm and anti-charm production.\\
This phenomena has been observed in hadron hadron collisions
\cite{e769a} and is well known as  
``leading effect". It has been the subject of many models
of particle production and several mechanisms have been proposed 
to explain it \cite{e687,ram-brod,cuautle}.
Here we study the $x_F$ distribution of $D^{\pm}$ and $D^0$ mesons 
produced in photon-proton collisions in the framework of a two-components
model that has been used  before to successfully describe the asymmetry in
pion proton interactions \cite{cuautle}.\\

The production of $D$ mesons in the model is assumed to take place {\it 
via} two different processes, namely QCD parton fusion with the 
subsequent fragmentation
of quarks in the final state and conventional recombination of valence
and sea quarks present in a vector meson fluctuation of the photon.\\
The asymmetries obtained with the conventional soft charm component
as well as  with a hard charm component in the photon, are presented. 
We compare both with the experimental data available.

To quantify the difference in the production of charm and anti-charm mesons
an asymmetry $A$ is defined as in \cite{e687},

\begin{equation}
A(x_F,p_t)=\frac{N_c - N_{\bar c}}{N_c + N_{\bar c}}.
\label{asym}
\end{equation}

\noindent
where $N_c$ and $N_{{\bar c}}$ are the production yields.
The asymmetry has been observed to be a function of both $x_F$ and the
transverse momentum $p_t$.\\

This paper is organized as follows:
the photon gluon mechanism for charm production will be discussed in
section 2. This mechanism is not responsible of a production asymmetry between
charm and anti-charm.
In section 3 we discuss the contribution of the resolved photon 
in the frame of a Vector Dominance Model (VDM) component. In subsection
3.1 we calculate the QCD interaction of the resolved photon. In
subsection 3.2, a recombination mechanism is discussed.
While the QCD production mechanisms in photoproduction are the same for 
charm and anti-charm production, recombination favors the formation of 
$D^-$ over $D^+$ and $\bar{D^0}$ over $D^0$.
In section 4 the various components are put together and the
asymmetry is estimated. Finally some conclusions are drawn in section 5.

\section{The photon-gluon fusion mechanism}

In this section we outline the calculation of the photon-gluon
fusion contribution at Leading Order (LO) in pQCD to the D-meson    
inclusive $x_F$ distribution.

The processes involved in the photoproduction of charm at LO in pQCD
are depicted in Fig. 1. The corresponding formula in the
parton model is
\begin{equation}
E_cE_{\bar{c}} \frac{d\sigma}{d^3p_c d^3p_{\bar{c}}} =
\int_0^1{dx \; g(x,Q^2) E_cE_{\bar{c}} \frac{d\hat{\sigma}}
{d^3p_c d^3p_{\bar{c}}}}
\label{pg1}
\end{equation}
where $g(x,Q^2)$ is the gluon probability density in the proton and
\begin{equation}
E_cE_{\bar{c}} \frac{d\hat{\sigma}}{d^3p_c d^3p_{\bar{c}}} =
\frac{1}{2\hat{s}} \frac{\alpha_{e}\alpha_{s}(Q^2)}{4(2\pi)^6}
(2\pi)^4\delta(p_{\gamma}+p_{g}-p_c-p_{\bar{c}})\left|{\it M}\right|^2 \;
.
\label{pg2}
\end{equation}

The squared invariant matrix $\left|{\it M}\right|^2$ in terms of the
Mandelstam variables is given by (see e.g. \cite{fontannaz,jones})
\begin{equation}
\left|{\it M}\right|^2 =
\frac{8}{9} \left[ \frac{1}{2} \left( 
\frac{m_c^2-\hat{t}}{m_c^2-\hat{u}} \right.
+ \frac{m_c^2-\hat{u}}{m_c^2-\hat{t}} \right) +
2\left( \frac{m_c^2}{m_c^2-\hat{t}} + 
\frac{m_c^2}{m_c^2-\hat{u}} \right)
-2\left( \frac{m_c^2}{m_c^2-\hat{t}} +
\left. \frac{m_c^2}{m_c^2-\hat{u}}\right)^2\right]
\label{pg3}
\end{equation}
and
\begin{eqnarray}
\hat{s} & = & xs  \nonumber \\
\hat{t} & = & m_c^2 - m_T\sqrt{s}e^{-y_c} \nonumber \\
\hat{u} & = & m_c^2 - xm_T\sqrt{s}e^{y_c} \; , 
\label{pg4}
\end{eqnarray}
where $s$ is the c.m energy of the $\gamma-p$ system,
$m_T^2=m_c^2+p_T^2$, $x$ is the momentum fraction of the gluon
inside the proton and $y_c=(1/2)\; ln[(E_c-p_c)/(E_c+p_c)]$ the rapidity
of
the $c$-quark.

After integrating out Eq. \ref{pg1} on the $\bar{c}$-quark variables
and the momentum fraction $x$, the inclusive cross section for the
production of charm (anti-charm) is given by   
\begin{equation}
\frac{d\sigma}{dx_F} = \frac{\sqrt{s}}{2}
\int{dp_T^2 \frac{xg(x,Q^2)}{E(\sqrt{s}-m_Te^{y_c})} 
\frac{d\hat{\sigma}}{d\hat{t}}}
\label{pg5}
\end{equation}
with
\begin{eqnarray}
\frac{d\hat{\sigma}}{d\hat{t}} & = & \frac{\pi\alpha_{e}\alpha_{s}(Q^2)}
{\hat{s}^2}\left|{\it M}\right|^2 \nonumber \\ 
x & = & \frac{m_Te^{-y_c}}{\sqrt{s}-m_Te^{y_c}} \; .
\label{pg6}
\end{eqnarray}
In Eq. \ref{pg6}, $\alpha_{s}$ is given by 
\begin{equation}
\alpha_{s}(Q^2) = \frac{12\pi}{(33-2N_f) log{\frac{Q^2}{\Lambda^2}}}
\label{pg7}
\end{equation}
with $N_f = 4$ and $\Lambda^2 = \Lambda_4^2$ according to the
gluon distribution used in Eq. \ref{pg5}.

In our calculations we use the GRV-LO gluon distribution \cite{gr-zpc}
with $Q^2=4m_c^2$, $m_c = 1.5$ GeV.
The D-meson inclusive $x_F$ distribution including fragmentation is
given by
\begin{eqnarray}
\frac{d\sigma_D}{dx_F} & = & \int{\frac{dz}{z}\;
\frac{d\sigma_{c(\bar{c})}}{dx} D_{D/c}(z)} \nonumber \\
z & = & \frac{x_F}{x} \; .
\label{pg8}
\end{eqnarray}
We use the Peterson fragmentation function,\\

\begin{equation}
D_{D/c}(z) =\frac{N}{z\left[1-\frac{1}{z} - \frac{\epsilon}{1-z} \right]^2}. 
\end{equation}

Inclusion of Next to Leading Order (NLO) contributions into the
D-meson cross section does not produce appreciable changes neither in the
form of the D-meson distribution or in its normalization (see e.g.
\cite{ellis-nason}). At NLO the cross section for the
production of an anti-quark differs from the cross section for the
production of a quark, but  this effect is small \cite{ellis-nason}.

\section{Light quark corrections to charm photoproduction}

We will identify the hadron like part of the photon with
$\rho$ and $\omega$ vector mesons neglecting the contribution of heavier
resonances which have smaller couplings to the photon.
The $\rho$ and the $\omega$ can be regarded as two states systems
\begin{eqnarray}
\rho ^0 = \frac{1}{\sqrt{2}} \left( u{\bar u} +  d {\bar d} \right); &
\omega = \frac{1}{\sqrt{2}} \left( u{\bar u} -  d {\bar d} \right).\\
\end{eqnarray}
The photoproduction of D mesons may take
place by QCD interaction of the vector meson $V$ with the proton and/or
by recombination of its constituent quarks (see Fig. 2).
Therefore, in obtaining the differential cross section of the process
$V p \rightarrow D X $, we will consider two possible
processes,
\begin{equation}
\frac{d \sigma}{d x_F}(V p \rightarrow D X) = 
\frac{d \sigma^{qcd}_{VDM}}{d x_F} (V p \rightarrow D X) + 
\frac{d \sigma^{rec}_{VDM}}{d x_F} (V p \rightarrow D X).
\end{equation}

In order to calculate these two contributions to the total cross section,
we assume that the momenta distribution of the quarks of a $\rho, \omega$
meson are the same than inside a pion. We will use the GRV parametrization for the parton distribution in the pion.\\

\subsection{QCD resolved photon contribution}

The photon may interact  through ist constituents  with the partons in the
proton.
In the parton fusion mechanism, $D^{\pm}$ $D^0(\bar{D}^0)$  mesons could
be produced via the $q\bar{q}(gg) \rightarrow c \bar{c}$ with the 
subsequent fragmentation of the $c(\bar{c})$ quark.
The contribution of the hadronic (or resolved)
component of the photon is given by the usual formula
that describes hadron-hadron interactions,
\begin{equation}
\frac{d\sigma^{qcd}_{VDM}}{dx_F} =
\frac{\sqrt{s}}{2} \sum_{i,j} \int{dp_T^2dy_4
\frac{x_1f_i(x_1,\mu)x_2f_j(x_2,\mu)}{E} \frac{d\hat{\sigma}}{d\hat{t}}}
\frac{D_{D/c} \left( z \right)}{z}
\label{pf1}
\end{equation}
where $x_1f_i(x_1,\mu^2)$ is the parton distribution in the resolved
photon,
$x_2f_i(x_2,\mu^2)$ is the parton distribution in the proton, $E$ is the
energy of the fragmenting charm quark and $D_{D/c}(z)$ is the
fragmentation function.

The partonic cross section in Eq. \ref{pf1} is given by
\begin{equation}
\frac{d\hat{\sigma}}{d\hat{t}} =
\frac{\pi\alpha_s^2(\mu)}{\hat{s}^2}
\left[\overline{\sum}\mid M_{i,j}\mid^2_{q\bar{q}} +
\overline{\sum}\mid M_{i,j}\mid^2_{gg}\right]
\label{pf2}
\end{equation}
with the invariant matrix elements squared and averaged (summed)
over initial (final) colours and spins at LO in pQCD given by
\begin{eqnarray}
\overline{\sum_{q\bar{q} \rightarrow c\bar{c}}}\mid M_{i,j}\mid^2 &=&
\frac{4}{9}\frac{\left(\hat{t} - m_c^2\right)^2 + \left(\hat{u} -
m_c^2\right)^2 + 2m_c^2\hat{s}}{\hat{s}^2} \nonumber \\
\overline{\sum_{gg \rightarrow c\bar{c}}}\mid M_{i,j}\mid^2 &=&
12 \frac{\left(m_c^2 - \hat{t}\right)\left(m_c^2 - \hat{u}\right)}
{\hat{s}^2}+\frac{8}{3} \frac{\left(m_c^2 - \hat{t}\right)
\left(m_c^2 - \hat{u}\right) - 2m_c^2\left(m_c^2 +
\hat{t}\right)}{\left(m_c^2 - \hat{t}\right)^2}  \nonumber \\
&+&\frac{8}{3} \frac{\left(m_c^2 - \hat{t}\right)\left(m_c^2 -
\hat{u}\right) - 2m_c^2\left(m_c^2 + \hat{u}\right)}{\left(m_c^2 -
\hat{u}\right)^2}-\frac{2}{3}\frac{m_c^2\left(\hat{s} -
4m_c^2\right)}{\left(m_c^2 -
\hat{t}\right)\left(m_c^2 - \hat{u}\right)}  \nonumber \\
&-&6  \frac{\left(m_c^2 - \hat{t}\right)\left(m_c^2 - \hat{u}\right) +
m_c^2\left(\hat{u} - \hat{t}\right)}{\hat{s}\left(m_c^2 - \hat{t}^2
\right)} \nonumber \\
&-&6 \frac{\left(m_c^2 - \hat{t}\right)\left(m_c^2 - \hat{u}\right)
+ m_c^2\left(\hat{t} - \hat{u}\right)}{\hat{s}\left(m_c^2 -
\hat{u}^2\right)}.
\label{pf3}
\end{eqnarray}
Writing the four momentum of the incoming and outgoing particles as
\begin{eqnarray}
p_1 & = & \frac{\sqrt{S}}{2}\left( x_1,0,0,x_1 \right) \nonumber \\   
p_2 & = & \frac{\sqrt{S}}{2}\left( x_2,0,0, -x_2 \right) \nonumber \\
p_c & = & \left( m_T \cosh(y_c),p_T,0,m_T \sinh(y_c)\right) \nonumber \\
p_{\bar{c}} & = & \left( m_T \cosh(y_{\bar{c}}),-p_T,0,m_T
\sinh(y_{\bar{c}})\right) \; ,
\label{pf4}
\end{eqnarray}
the Mandelstam variables appearing in Eqs. \ref{pf3} are given by
\begin{eqnarray}
\hat{s} & = & 2m_T^2 \left( 1+\cosh(\Delta y)\right) \nonumber \\
\hat{t} & = & m_c^2 - m_T^2 \left(
1+\exp(-\Delta y)\right) \nonumber \\
\hat{u} & = & m_c^2 - m_T^2 \left( 1+\exp(\Delta y)\right) \\
\Delta y & = & y_c - y_{\bar{c}}. \nonumber
\label{pf5}
\end{eqnarray}
In our calculation we use the GRV-LO \cite{gr-zpc} parton distributions in
protons and pions, and apply a global factor of $2-3$
in order to account for NLO effects. For the fragmentation function we
use the Peterson function.

\subsection{Charmed meson production by recombination}

In the scenario described in \cite{bednyakov}
for $\pi^-$ proton collisions, the annihilation of a $u$ quark
from the proton and the $\bar{u}$ quark in the pion would liberate the $d$
of the pion which in turn recombines to form a $D^-$ and certainly not a
$D^+$. 
On a similar base a $\rho^0$ proton collision will favor the production of
$\bar{D^0}$ and $D^-$ over $D^0$ and $D^+$ depending on the quantum state
of the colliding $\rho^0$ at the interaction point (see Fig. 3).\\

The production of leading mesons at low $p_T$ was described by 
recombination of quarks long time ago \cite{dh-plb}.
In recombination models one assumes that the outgoing 
hadron is produced in the beam fragmentation region through the 
recombination of the maximum number of valence quarks and the minimum 
number of sea quarks of the incoming hadron.
The invariant inclusive $x_F$ distribution for leading mesons is given by
\begin{equation} 
\frac{2 E}{\sqrt {s}\sigma}\frac{d\sigma^{rec}}{dx_F}=
\int_0^{x_F}\frac{dx_1}{x_1}\frac{dx_2}{x_2}
F_2\left( x_1,x_2\right) 
R_2\left( x_1,x_2,x_F\right) 
\label{rec-cs}
\end{equation}
where $x_1$, $x_2$ are the momentum fractions and
$F_2 \left( x_1,x_2\right) $ is the two-quark distribution
function of the incident hadron. $R_2 (x_1,x_2,x_F)$ is the 
two-quark recombination function. \\
The two-quark distribution function is parametrized in terms of  
the single quark distributions.
For recombination of $D^-$, $D^0$
\begin{equation}
F_2 \left( x_1,x_2 \right) = 
\beta F_{d,u;val}\left(x_1\right)F_{\bar{c};sea}\left(x_2\right)
\left(1-x_1-x_2\right),
\label{3-quark}
\end{equation}
with $F_{q}\left(x_i\right) = x_iq\left(x_i\right)$.
We use the GRV-LO parametrization for the single quark distributions in
Eq. \ref{3-quark}.
 It must be noted that since the 
GRV-LO \cite{gr-zpc} distributions are functions of $x$ and $Q^2$, 
our  $F_2 \left( x_1,x_2 \right)$ also depends on $Q^2$.\\
The recombination function is given by
\begin{equation}
R_2 \left( x_{u,d},x_{ \bar{c} }\right) =\alpha 
\frac{x_{u,d} x_{\bar{c}} }{x_F^2}
\delta \left(x_{u,d}+x_{\bar{c}}-x_F\right) \: ,   
\label{eq7}
\end{equation}
with $\alpha$ fixed by the condition 
$\int_0^1 dx_F (1/\sigma)d\sigma^{rec}/dx_F = 1$.\\
Some time ago V. Barger {\it et al.} \cite{halzen} explained the
spectrum enhancement at high $x_F$ in $\Lambda_c$  production 
assuming a hard momentum distribution of charm in the proton.
Here we will also take  a QCD evolved charm distribution, of
the form proposed by V. Barger {\it et al.} \cite{halzen}
\begin{equation}
xc(x, \langle Q^2 \rangle ) = N x^l (1-x)^k,
\label{barger}
\end{equation}
with a normalization $N$ fixed to 
\begin{equation}
\int dx \cdot x c(x) = 0.005
\end{equation}
and $l=k=1$. With this values for $l$ and $k$ one tries to resemble
the distribution of valence quarks. In contrast with the parton fusion
calculation, in which the scale  $Q^2$ of the interaction is fixed at the
vertices of the appropriated Feynman diagrams, in recombination the value
of the parameter $Q^2$ should be used to give adequately the content of
the recombining quarks in the initial hadron. We used $Q^2=4 m_c^2$.\\

\section{$D^{\pm}$ and $D^0(\bar{D}^0)$ total production}

The inclusive production cross section of a $D$ meson is then
obtained by adding the contribution of recombination, Eq. \ref{rec-cs}, 
to the QCD processes from direct photon-gluon interaction, quark
anti-quark annihilation and gluon-gluon fusion from the hadronic
component of the photon, {\it i. e.}
\begin{equation}
\frac{d\sigma^{tot}(D^-)}{dx_F} =N_1 \left( \frac{d\sigma^{\gamma g} }{dx_F} + 
a \left(
b \frac{d\sigma^{qcd}_{VDM}}{dx_F} + 
c \frac{d\sigma^{rec}_{VDM}}{dx_F} \right) \right)
\label{sig-tot1}
\end{equation}
\begin{equation}
\frac{d\sigma^{tot}(D^+)}{dx_F} = N_2\left( \frac{d\sigma^{\gamma g} }{dx_F} + 
a \left(
b \frac{d\sigma^{qcd}_{VDM}}{dx_F} + 
d \frac{d\sigma^{rec}_{VDM}}{dx_F} \right) \right)
\label{sig-tot2}
\end{equation}
with
\begin{equation}
\frac{d\sigma^{qcd}_{VDM}}{dx_F} = \frac{d\sigma^{g g} }{dx_F} + 
\frac{d\sigma^{q\bar{q}}}{dx_F}
\end{equation}
\noindent
and  $N_1= \frac{1}{1+ab+ac}$, $N_2 =\frac{1}{1+ab+ad}$ the parameters $a,b,c$
and $d$ depend on the contribution of each process to the total cross 
section. They are fixed in such a way that the differential cross section is
well described, before calculating the asymmetry.
The resulting inclusive $D$ production cross section
$d\sigma^{tot}/dx_F$ (shown in Fig. 4), is used then to construct the 
asymmetry defined in Eq. \ref{asym}.\\
The values obtained for the different contributions in Eqs. 24 and 25  are 
in reasonable  agreement with what one would expect. The photon fluctuation
to a vector meson is of the oder of 1 $\%$. Approximately 96 $\%$ of the total 
cross section comes from the photon gluon process. The contribution due to 
recombination goes from about 1 $\%$ (for $D^-$) to about 3 $\%$ (for $D^+$). 
Fig. 5 shows the model prediction for the $D^-$, $D^+$ production 
asymmetry together with the experimental results from the E687 Collaboration
\cite{e687}. The two curves correspond to the conventional GRV function 
distribution in the resolved photon and to the distribution proposed by Barger, 
{\it et al.} where a hard charm component  has been assumed.

\section{Conclusions}

In an earlier work \cite{nos1} the production asymmetry of $\Lambda_c$ was
described using the same recombination scheme used here. In hadroproduction 
the presence of a diquark in the initial state, plays an important 
role in $\Lambda_c$ production. In photo production, however, the production 
mechanism is somewhat different and the asymmetry is much smaller.
The parameters used here are in reasonable agreement with what is physically 
expected and with the values used in a previous study of production 
asymmetries in $\pi$ proton collisions \cite{cuautle}. Changing the values of
these parameters may improve the description of the asymmetry but then they 
may loose meaning in the frame of the asymmetry obtained for hadro production.\\

\noindent
Other experimental results for the asymmetry defined as:\\
\begin{equation}
A(x_F) = \frac{\sigma(D^+) - \sigma(D^-)}{\sigma(D^+) + \sigma(D^-) }
\end{equation}
are $A= -0.0384 \pm 0.0096$ in  $x_F \ge 0.0$ at photon energies of 200 GeV
\cite{e687} and 
$A= -0.0196 \pm 0.0147$ in  $x_F \ge 0.2$ at photon energies of 80-230
GeV, average energy 145 GeV \cite{e691}. The NA14 collaboration studied
the asymmetry
\begin{equation}
A(x_F) = \frac{\sigma(D^+ + D^0) - \sigma(D^- + \bar{D^0})}
{\sigma(D^+ + D^0) + \sigma(D^- + \bar{D^0}) }
\end{equation}
and obtained $A= -0.03 \pm 0.05$   $x_F \ge 0.0$ at photon energies of 40-140
GeV and 100 GeV  in average \cite{na14}.\\
All these results are in good agreement with each other
but, the statistical errors are still large.

In ref.~\cite{e687} experimental data are compared to a model based on string 
fragmentation. This model gives a larger asymmetry than the one obtained in our
approach. The description obtained there is in much better agreement with the
experimental results. However, more precise measurements are needed before one 
can draw a final answer.\\

One would expect that with increasing energy the resolved photon
component increases giving rise to a larger contribution of the 
recombination mechanism which in turn would produce a larger production
asymmetry. The HERA experiments should therefore be able to see a production 
asymmetry. 
However larger energies of the photon means also 
smaller values of $x$ for the quarks that participate in the interaction
and the hard charm component is expected to play a minor role
at small $x$'s. {\it i e.}, the asymmetry should look rather as the one 
obtained from conventional densities. New experiments will have 
soon more precise measurements of the asymmetries. This new results will
give us a better understanding of the underlying production mechanisms.


\newpage

\section*{Figure Captions}
\begin{itemize}
\item
[Fig. 1:] According with perturbative QCD, photon-gluon fusion is
the main process in charm photo-production.
\item
[Fig. 2:] The resolved photon may interact with the proton via QCD, i.e.
quark-antiquark annihilation and gluon-gluon fusion as shown in (a,b).
A non perturbative interaction (c) may favor charm  over
anti-charm mesons production.
\item 
[Fig. 3:] After a fluctuation of the photon to a $\rho ^0$ vector meson,
the interaction with the proton may ocurr in one of the two states.
In any case the valence quark in the $\rho$ would be released once
the antiquark annihilates with a quark of the proton.
\item 
[Fig. 4:] Total cross section as a function of $x_F$. Experimental results 
from ~\cite{e691} and theoretical calculation as in Eqs. \ref{sig-tot1} and 
\ref{sig-tot2} using the GRV distributions.
\item
[Fig. 5:] Measured production asymmetry for $D^-$ and $D^+$ from [1]. 
The curves show the model result in which a hard charm  component 
(dashed line) and GRV-LO (solid line) in the pion has been 
considered. The horizontal line at $A(x_F)=0$ is for reference only.
\end{itemize}

\newpage
\begin{figure}[b]
\psfig{figure=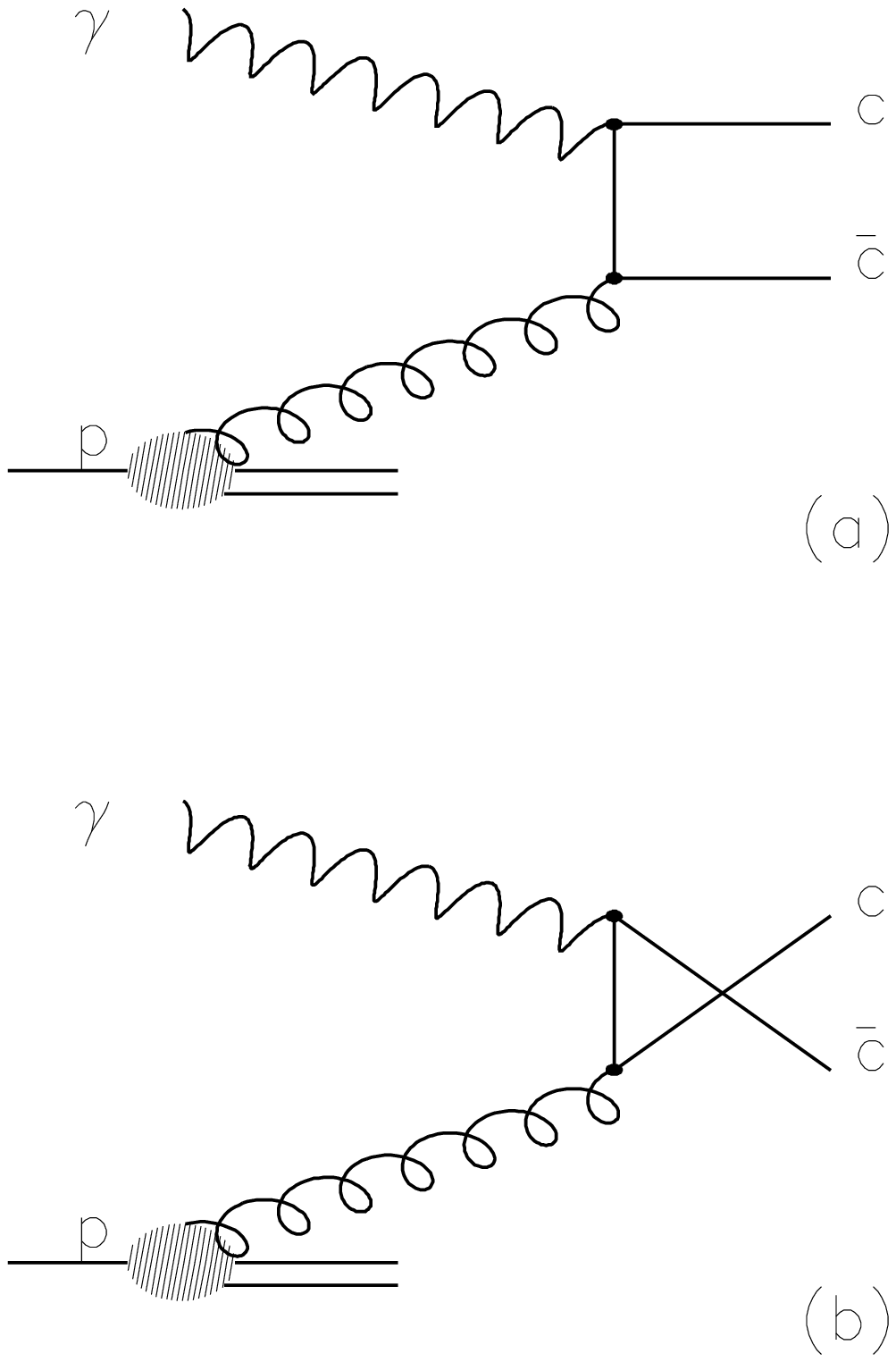,height=6.0in}
\caption{}
\end{figure}
\begin{figure}[b]
\psfig{figure=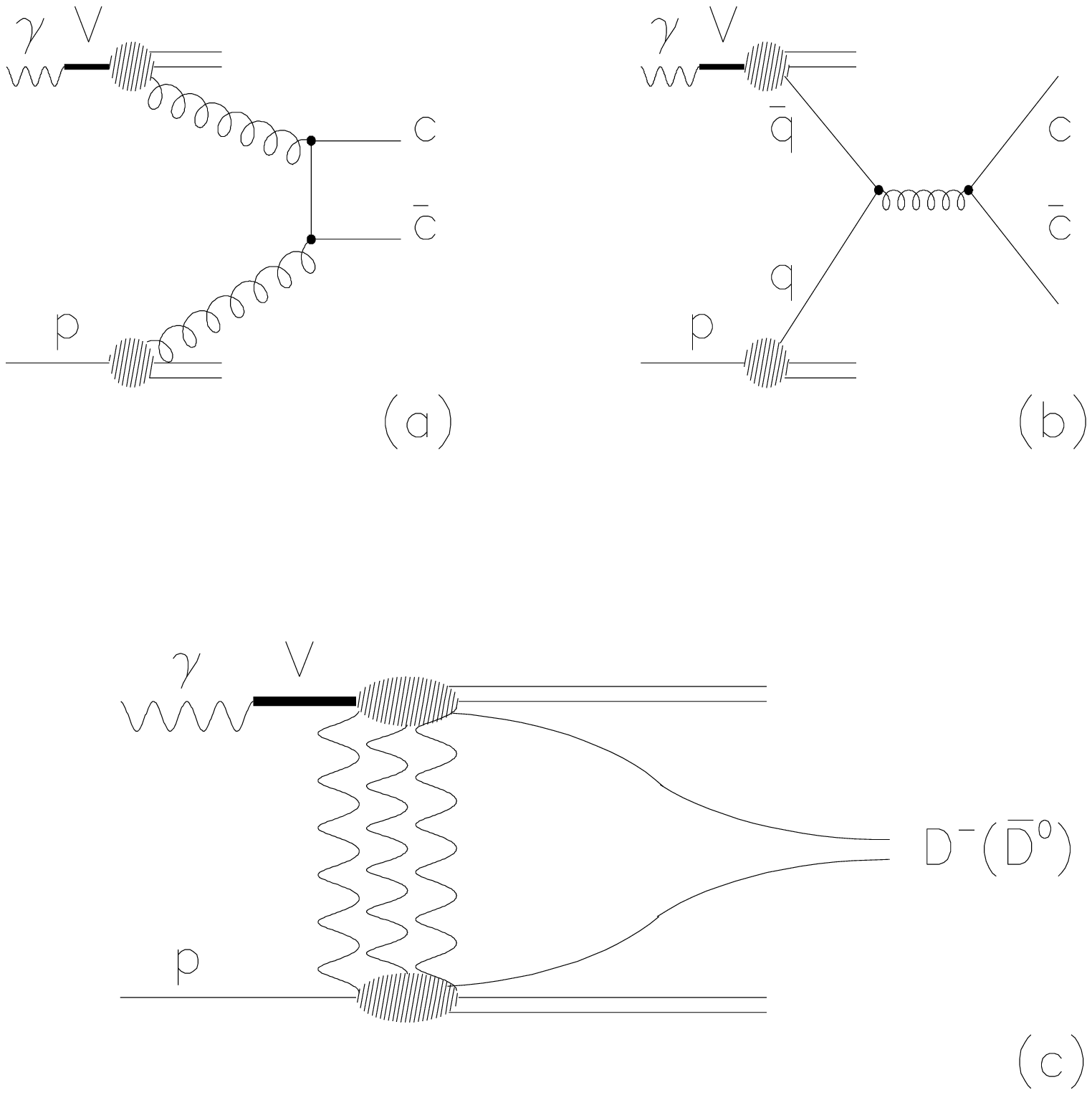,height=6.0in}
\caption{}
\end{figure}
\begin{figure}[b]
\psfig{figure=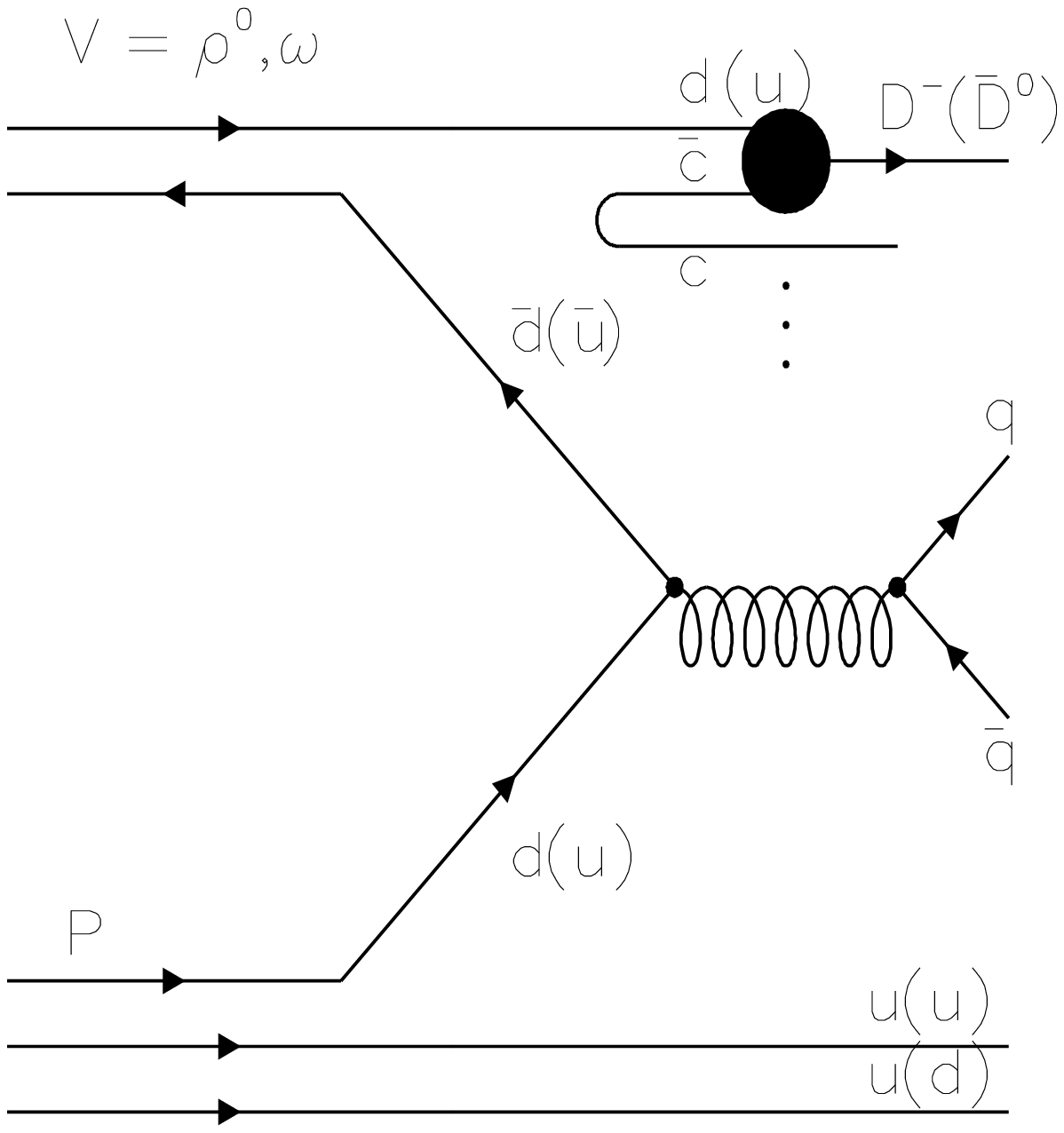,height=6.0in}
\caption{}
\end{figure}

\begin{figure}[b]
\psfig{figure=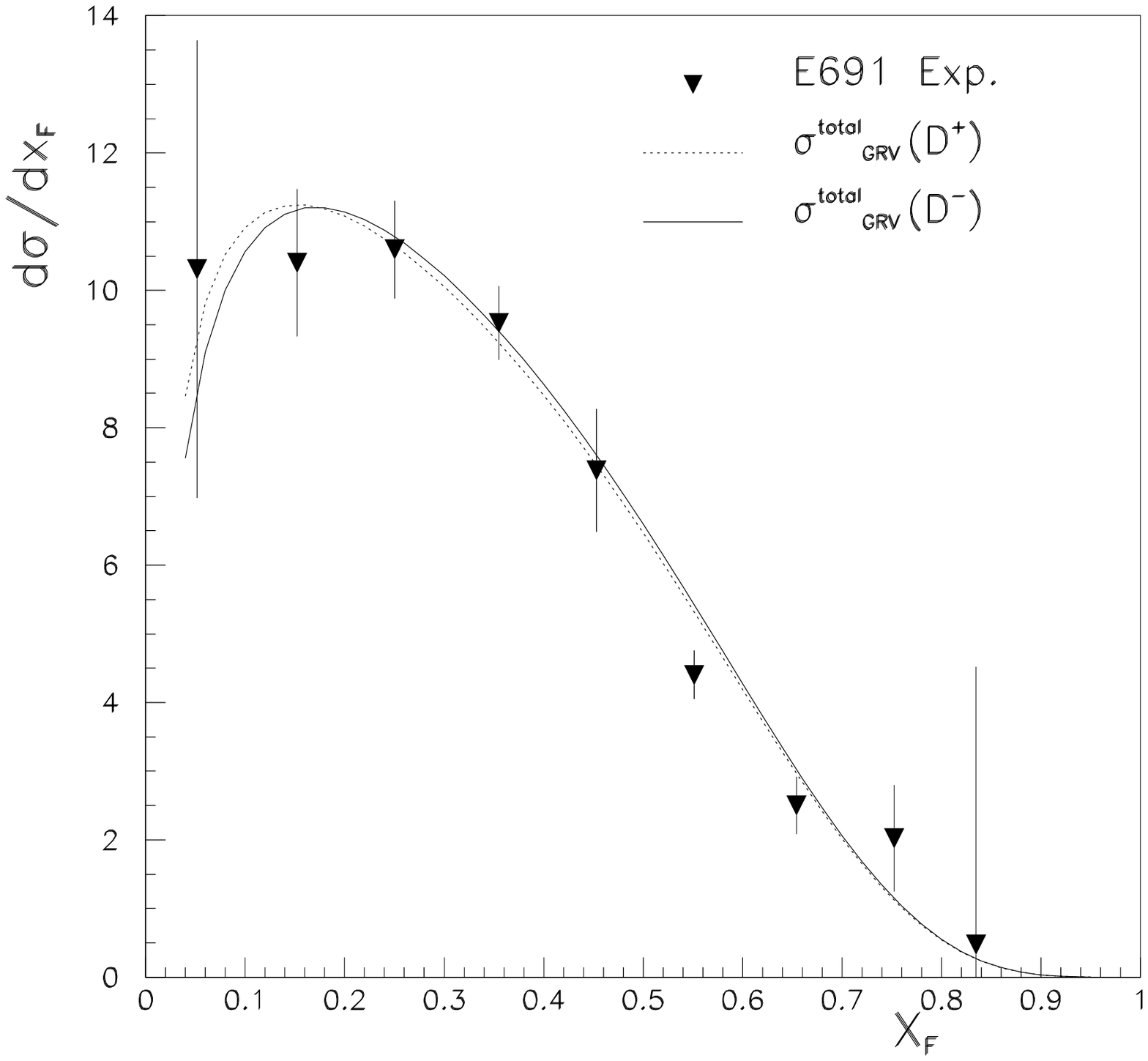,height=6.0in}
\caption{}
\end{figure}
\begin{figure}[b]
\psfig{figure=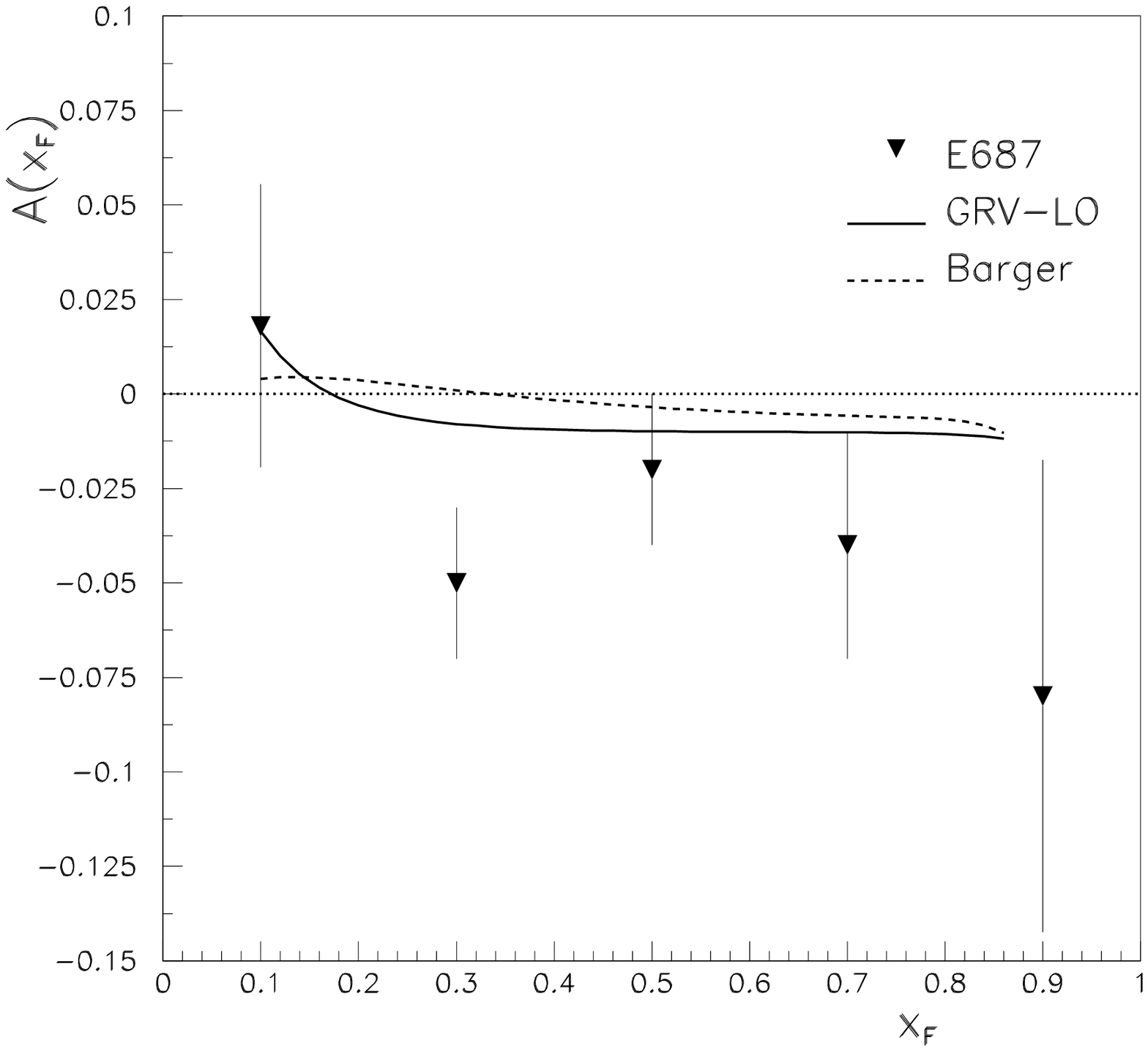,height=6.0in}
\caption{}
\end{figure}

\end{document}